\documentclass[prc,twocolumn,showpacs,showkeys,preprintnumbers,floatfix,nofootinbib,
superscriptaddress,longbibliography]{revtex4-2}

\usepackage{amsfonts} % AMS
\usepackage{amssymb} % AMS
\usepackage{amsmath} % AMS
\usepackage{graphicx} % Include figure files
\usepackage{subfigure} % Include figure files
\usepackage{array} % array
\usepackage{dcolumn} % Align table columns on decimal point
\usepackage{bm} % bold math
\usepackage{latexsym} % latex symbols
\usepackage{longtable} % long tables
\usepackage{bbold}
\usepackage[colorlinks=true, linkcolor = blue, citecolor=blue,urlcolor=blue, bookmarksnumbered=true, bookmarksopen=true]{hyperref} % hypertext links 

\usepackage[qm,braket]{qcircuit}

\usepackage[dvipsnames,usenames]{xcolor}
\usepackage{xcolor}
\definecolor{vermillion}{rgb}{0.86, 0.18, 0.01}

\begin{document}

%%%

\preprint{LA-UR-21-30446, FERMILAB-PUB-21-493-QIS, INT-PUB-21-022, IQuS@UW-21-011}

%\title{Efficient excited state preparation for linear %response}
\title{Nuclear two point correlation functions on a quantum computer}

%% author list might be incomplete / wrong
%% simply put in alphabetical order (might be wrong/incomplete, too)

\author{A.~Baroni}
%\email{}
\affiliation{Theoretical Division, Los Alamos National Laboratory, Los Alamos, NM 87545, USA}

\author{J.~Carlson}
\affiliation{Theoretical Division, Los Alamos National Laboratory, Los Alamos, NM 87545, USA}

\author{R.~Gupta}
\affiliation{Theoretical Division, Los Alamos National Laboratory, Los Alamos, NM 87545, USA}

\author{Andy C.~Y.~Li}
\affiliation{Fermi National Accelerator Laboratory, Batavia, IL, 60510, USA}

\author{G.~N.~Perdue}
\affiliation{Fermi National Accelerator Laboratory, Batavia, IL, 60510, USA}

\author{A.~Roggero}
\affiliation{Physics Department, University of Trento, Via Sommarive 14, I-38123 Trento, Italy}
\affiliation{INFN-TIFPA Trento Institute of Fundamental Physics and Applications,  Trento, Italy}
\affiliation{InQubator for Quantum Simulation (IQuS), Department of Physics, University of Washington, Seattle, WA 98195, USA}

\begin{abstract}
The calculation of dynamic response functions is expected to be an early application benefiting from rapidly developing quantum hardware resources. The ability to calculate real-time quantities of strongly-correlated quantum systems is one of the most exciting
applications that can easily reach beyond the capabilities of traditional classical hardware. Response functions of fermionic
systems at moderate momenta and energies corresponding roughly to the Fermi energy of the system are a potential early application because the relevant operators are 
nearly local and the energies can be resolved in moderately short real time, reducing the
spatial resolution and gate depth required.

This is particularly the case in quasielastic electron and neutrino scattering from nuclei, a topic of great interest in the nuclear and particle physics communities and directly related to experiments designed to probe neutrino properties. In this work we use current quantum hardware and error mitigation protocols to calculate response functions for a highly simplified nuclear model through calculations of a 2-point real time correlation function for a modified Fermi-Hubbard model in two dimensions with three distinguishable nucleons on four lattice sites.
\end{abstract}

\maketitle

%\title{Efficient excited state preparation for linear %response}

\section{Introduction}

Quantum computing holds the promise of enabling calculations of the real-time evolution of quantum systems, with a wide range of applications across many areas of physics including electronic many-body problems, condensed matter, cold atom, nuclear and particle physics.
Quantum dynamics with more than a few particles easily exceeds the capabilities of traditional computers because of the extremely large number of basis states and the oscillatory nature of the path integrals involved in evaluating cross sections and transition rates. In specific cases such
as low-energy resonance scattering and high-energy semi-classical approaches, valuable information can be gained with
classical computers~\cite{Pastore2018,Pastore2020}.
Quantum computing, though, has the potential to perform calculations of quantum dynamics beyond the reach of classical computing.

The linear response of quantum systems is
%, in many cases,
a promising candidate for early applications of quantum computers~\cite{Baez2020}.
Linear response, as measured for example in neutron scattering from materials or electron and neutrino scattering from nuclei directly probes the structure and dynamics of the underlying system.  By adjusting the momentum and energy transfer one can focus on different scales. Even seemingly simple cases, where the transfers are of the order of the Fermi momentum, can yield rich physics. While at larger momentum transfers the response is largely a function of the momentum distribution or spectral function of the target\cite{Benhar:1989,Rocco:2019}, at more modest momenta two-nucleon physics including two-nucleon currents, correlations, and charge exchange
can become quite important~\cite{Lovato2016,Pastore2020,Lovato2020,Gysbers_2019}. These effects
can play a role, for example, when trying to measure neutrino properties through neutrino-nucleus scattering in experiments including MiniBooNE, MicroBooNE, T2K and DUNE~\cite{Lovato2020}.

The linear response function for a specific quantum state (often the ground state) is defined as:
\begin{equation}
\label{eq:response}
    S(\omega,{\bf q}) = \sum_f |\langle \Psi_0 | O ({\bf q}) | f \rangle|^2 \delta (\omega - (E_f - E_0))\;,
\end{equation}
% S(\omega,{\bf q}) = \sum_f \langle \Psi | O ({\bf q}) | f \rangle \langle f | O ({\bf q}) | \Psi \rangle \delta (\omega - (E_f - E_0)),\nonumber\\
%\end{eqnarray}
where $\ket{\Psi_0}$ describes the ground state
with energy $E_0$,
the sum includes all final states $\ket{f}$ with energy $E_f$, $\omega$ and ${\bf q}$  denote respectively the energy and the three-momentum transfer injected in the system by the external probe, and the
operator $O({\bf q})$ describes the coupling of the system to the external probe.  For electron scattering from nuclei there are two response functions describing the longitudinal (charge) and transverse (current) components of the response~\cite{Carlson98}.  For neutrino scattering there are five response functions for a given
momentum and energy transfer, the cross section is given by a linear combination of these response functions that depend upon geometry (e.g. the lepton scattering angle) and the particle (e.g. electron, neutrino, or anti-neutrino) being scattered~\cite{Shen2012}.
The response functions govern the inclusive cross section where all final scattering states are summed over and only the lepton kinematics is specified.

Equivalent information is available from the Fourier transform of 
the relativistic two-point correlation function in real time $\tau$:
\begin{equation}
    S_M(q_0,{\bf q}) = \int d^4 x e^{iqx} \bra{\Psi_0} O^\dagger({\tau,\bf x}) O({0,0}) \ket{\Psi_0} \,,
%    C(\tau,{\bf q}) = \bra{\Psi_0} O^\dagger({\bf q}) \exp [ -i H \tau] O({\bf q}]) \ket{\Psi_0} .
\end{equation}
which in the non-relativistic limit is the same as the real-time equivalent of  Eq.~\eqref{eq:response}, ie, $S_M(q_0,{\bf q}) \to S(\omega,{\bf q})$. 
Classical simulations in discrete Euclidean time $\tau_E$, for example of lattice QCD, give an analogous two-point correlation function:\looseness-1
\begin{equation}
    C_E(\tau_E,{\bf q}) = \sum_{\bf x} e^{i{\bf q}\cdot {\bf x}} \bra{\Psi_0} O^\dagger({\tau_E,\bf x})  O({0,0}) \ket{\Psi_0} ,
%    C(\tau,{\bf q}) = \bra{\Psi_0} O^\dagger({\bf q}) \exp [ -i H \tau] O({\bf q}]) \ket{\Psi_0} .
\end{equation}
where in both cases the operator $O$ can be the electromagnetic current. We recall that $C_E$ and $S_M$ are related through the following Laplace transform $C_E(\tau_E,{\bf q}) = \int d\tau_E e^{-q_0 \tau_E} S_M(q_0,{\bf q})$. The problem of obtaining $S_M$ from $C_E$ is in general an ill-posed problem. The lattice calculation can be done only at a finite number of 
discrete values of $\tau_E$ with the data having errors that typically grow very rapidly with $\tau_E$. These errors get exponentially amplified by the inversion procedure. Various techniques have been developed in order to obtain good approximations of the frequency response $S_M$ starting from euclidean data~\cite{Silver1990,Vitali2010,Burnier2013,Kades2020,Raghavan2021}. These approaches are often able to reconstruct signals with simple structure, such as a quasi elastic peaks which are composed of one broad peak~\cite{Lovato2016}, or observables that can be recast directly into euclidean time like transport coefficients dominated by zero frequency modes (see eg.~\cite{Roggero_2016}), but the systematic errors are difficult to assess reliably in the general case.

Quantum computers could perform simulations in real-time, however, a number of 
challenges needs to be solved such as the 
preparation of the ground state $\ket {\Psi_0}$, the application of the operator $O$, the evolution of the system 
for time $\tau$ long compared to the relevant energy scale, $1/q_0$, of the calculation while 
preserving coherence, and lastly transitioning back to the ground state through the second insertion of the excitation operator
$O$. This work is a step in that direction. 

In traditional approaches to the non-relativistic response, an integral transform of the response
is calculated (such as Laplace, Lorentz or Gaussian, see e.g.~\cite{Carlson:1992,Lovato2016,Roggero_2013,Efros_2007,Sobczyk:2021,Sobczyk_2021}). The
imaginary-time two point function is  
calculated in Quantum Monte Carlo approaches, and is equivalent to the non-relativistic limit
of lattice QCD evalutations.

At high energy and momenta, the response
can also be calculated through 
local properties of the initial state (momentum distribution or spectral function)~\cite{Benhar:1989,Rocco:2019}, short-time expansions of the nuclear propagation~\cite{Pastore2020}.
The single-particle Green's function yields information on this response, while the short-time approximation incorporates the
coupling and propagation of pairs of nucleons in the final state.  The short-time approach guarantees that the calculated response reproduces the energy independent and energy-weighted sum rules $E_0 (\bf q)$ and $E_{m > 0} ({\bf q})$ defined as:
\begin{equation}
E_m({\bf q}) = \bra{\Psi_0} O^\dagger({\bf q}) H^m O({\bf q}) \ket{\Psi_0},
\end{equation}
where $H^m$ is the Hamiltonian to the m${}^{\rm th}$ power, and is
available from a Taylor expansion of the real or imaginary-time response at short times.
Including the two-particle ladders in the propagation reproduces reasonably well  quasi-elastic scattering in simple test cases; it is equivalent to one sophisticated Trotter step in the real-time evolution as described in Ref.~\cite{Pastore2020}. This approach provides a natural explanation for the scaling with momentum and nuclear system size observed in electron scattering~\cite{Pastore2020}.
Quantum computers, though, can, in principle, follow the real-time evolution of the system over larger distances and longer times, enabling reconstruction of the response at lower momenta and energies.
They may also be used to calculate cross sections to explicit final states, a much more challenging task on even future quantum hardware~\cite{roggero2019}.

In this paper we study two-point functions of a simple nuclear model, first introduced in Ref.~\cite{roggero2020A}, using current quantum hardware, and recently developed error mitigation strategies ~\cite{kandala2017,Roggero2020b}.

\section{Lattice model and Response Function}
We follow the approach originally developed in Ref.~\cite{roggero2020A}, where we considered a model for the triton nucleus on a two-dimensional lattice with periodic boundary conditions, with two dynamical nucleons and one static particle on a specific lattice site.
The model considered is equivalent to a two species Hubbard model with two- and three- body interactions.
At the leading order in pion-less Effective Field Theory~\cite{Lee2004}, the Hamiltonian describing the low energy dynamics of a system of nucleons discretized on a lattice can be expressed as follows~\cite{roggero2020A,Lee2004,Borasoy2006}
\begin{equation}
\begin{split}
\label{eq:gen_model}
H=&-t\sum_{f=1}^{N_f}\sum_{\langle i,j\rangle}c^\dagger_{i,f} c_{j,f}+U\sum_i\sum_{f<f^\prime}^{N_f}n_{i,f}n_{i,f^\prime}\\
&+V\sum_{f<f^\prime<f^{\prime\prime}}\sum_i n_{i,f}n_{i,f^\prime} n_{i,f^{\prime\prime}}\, ,\\
%&+U\sum_{f=1}^{N_f}n_{1,f}+V\sum_{f<f^\prime}n_{1,f}n_{1,f^\prime}\, ,    
\end{split}
\end{equation}
where the particle number operator $n_{i,f}= c^\dagger_{if}c_{if}$, with $c_{if}$ the destruction operator at site $i$ of species $f$. The kinetic energy (or hopping term) in the first line contains a sum $\langle i,j\rangle$ over nearest-neighbor sites on the lattice. This corresponds to a generalized Hubbard model with $N_f$ fermionic species and the addition of an on-site repulsive three-body interaction. For applications in nuclear physics one typically considers $N_f=4$ fermionic species corresponding to neutrons and protons in both spin projections. The numerical values of the couplings $t$, $U$ and $V$ are reported in table 1 of Ref.~\cite{roggero2020A} (taken from \cite{Rokash2013}) and correspond to a lattice spacing of $1.4$ fm.

Similarly to Ref.~\cite{roggero2020A} we will also introduce an additional local term on site $1$ of the lattice given by
\begin{equation}
\begin{split}
\label{eq:on_site_int}
H_{site}=U\sum_{f=1}^{N_f}n_{1,f}+V\sum_{f<f^\prime}n_{1,f}n_{1,f^\prime}\, ,    
\end{split}
\end{equation}
with the same $V$ and $U$ used in the general model of Eq.~\eqref{eq:gen_model}.
As explained in Ref.~\cite{roggero2020A}, this additional interaction term mimics the presence of an additional particle at that spatial site that is treated statically.

%The Hamiltonian of interest, reported in  Eqs. (1)-(3) of Ref.~\cite{Roggero2020A}  and is reported below for completeness
%\begin{eqnarray}
%\label{eq:gen_model}
%H&=&-t\sum_{f=1}^{N_f}\sum_{\langle i,j\rangle}c^\dagger_{i,f} c_{j,f}+2dt\,A\nonumber\\
%&+&U\sum_i\sum_{f<f^\prime}^{N_f}n_{i,f}n_{i,f^\prime}+V\sum_{f<f^\prime<f^{\prime\prime}}\sum_i n_{i,f}n_{i,f^\prime} n_{i,f^{\prime\prime}}\nonumber\\
%&+&U\sum_{f=1}^{N_f}n_{1,f}+V\sum_{f<f^\prime}n_{1,f}n_{1,f^\prime}\, ,
%\end{eqnarray}
%where the particle number operator $n_{i,f}= c^\dagger_{if}c_{if}$ with https://www.overleaf.com/project/5e1901c13813510001243fdf$c_{if}$ the destruction operator at site $i$ of species $f$. The hopping term in the first line contains a sum $<i,j>$ over nearest-neighbor sites. 

%For a three dimensional lattice, the numerical values of the couplings $t$, $U$ and $V$ are reported in table 1 of Ref.~\cite{Roggero2020b} and correspond to a lattice of $1.4$ fm.
In order to describe scattering process with external probes, like neutrino-nucleus scattering, we now resort to linear response theory.
We consider an interaction of the system described by the Hamiltonian in Eq.~\eqref{eq:gen_model} with an external 'weak' probe injecting momentum 
\begin{eqnarray}
{\bf q}_{k}&=&\frac{\pi}{L}\left(x_k,y_k,z_k\right)\;,
\end{eqnarray}
with $L$ the spatial length of the lattice in 3 dimensions and $x_k,y_k,z_k$ positive integer numbers denoting the location of site $k$ on the reciprocal lattice. In this initial exploration we seek to describe probes that couple to the nucleon density as in neutron scattering or a part of the nuclear longitudinal response  and described by the following interacting Hamiltonian
\begin{equation}
\label{eq:int_ham}
H_I({\bf q}_{k})=\sum_{f=1}^{N_f}\rho_f({\bf q}_{i})=\sum_{f=1}^{N_f}e_f\sum_i e^{i{\bf q}_{k}\cdot{\bf r}_i}n_{i,f}\, ,   
\end{equation}
where ${\bf r}_i=\left(x_i,y_i,z_i\right)$ denotes the location of site $i$ on the spatial lattice, $\rho_f$ and $e_f$ denote respectively the charge density operator and the charge for the species $f$. We notice that from now on we will use the symbol $H_I$ to represent what we denoted with $O$ in Eq.~\eqref{eq:response} as the excitation operator.  We consider the system to be initially in its ground state $\ket{\Psi_0}$ and, in the linear response regime, the probability to transition to all the final states $\ket{n}$ can be obtained using Fermi's golden rule
\begin{eqnarray}
S(\omega,{\bf q}_k)&=& \sum_n  \left\lvert \bra{\Psi_0} H_I({\bf q}_{k})\ket{n}\right\rvert^2\delta(E_n+\omega-E_0)\, , 
\end{eqnarray}
analogue of Eq.~\eqref{eq:response}.
We recall now the known identity
\begin{equation}
\begin{split}
\label{eq:S_transform}
S(\omega,{\bf q}_{k})&=\int_{-\infty}^\infty \frac{d\tau}{2\pi} e^{i\omega \tau} \bra{\Psi_0}H_I(\tau,{\bf q}_{k})H_I({\bf q}_{k})\ket{\Psi_0}\\
&=\int_{-\infty}^\infty \frac{d\tau}{2\pi} e^{i\omega \tau} C(\tau,{\bf q}_{k})\, ,
\end{split}
\end{equation}
relating the response function in frequency to a two-point time correlation function $C(\tau,{\bf q}_{k})$. This definition uses $H_I(\tau,{\bf q}_{k})=e^{iH\tau}H_I({\bf q}_{k}) e^{-iH\tau}$ as the interacting Hamiltonian in the Heisenberg picture. One of the goals of the present work is to study the feasibility of reconstructing the frequency response $S(\omega,{\bf q}_{k})$ from a (possibly noisy) estimate of the correlation function $C(\tau,{\bf q}_{k})$ defined above. This is a classic problem in linear response theory and efficient quantum algorithms have been developed in the past for the calculation of the response function through both real-time simulations~\cite{Somma2002,OBrien2019,Somma2019} and by direct sampling in frequency space~\cite{roggero2019,roggero2020C}. Recent work also proposed quantum algorithms to study directly Green's functions~\cite{endo2020,Ciavarella_2020}. In this work, we estimate the two-point function $C(\tau,{\bf q}_{k})$ directly with a quantum simulation of a small system using current generation superconducting devices. We use an efficient algorithm originally proposed in Ref.~\cite{Somma2002} that requires one additional ancilla qubit and the possibility to apply the interaction Hamiltonian $H_I({\bf q}_{k})$ controlled on the ancilla state. This requires $H_I({\bf q}_{k})$ to be unitary, but the scheme can be easily generalized to interactions that admit a short expansion into unitaries as we do for our example model described in the next section. 

\subsection{Real time correlation functions}
The quantity that we will calculate in the following is the real time response function defined by the r.h.s. of Eq.~\eqref{eq:S_transform} explicitly reported below in a compact notation
\begin{eqnarray}
\label{eq:resp_real_time}
C(\tau,{\bf q}_k)&=&\bra{\Psi_0} U^\dagger(\tau)H_I({\bf q}_k)U(\tau)H_I({\bf q}_k)\ket{\Psi_0}\nonumber\\
&=&\langle U^\dagger(\tau)H_I({\bf q}_k)U(\tau)H_I({\bf q}_k)\rangle \, ,
\end{eqnarray}
with $U(\tau)=e^{-iH\tau}$ the real time evolution operator. 
We can express the interacting Hamiltonian, after a mapping to qubits, as a sum of Pauli operators
\begin{eqnarray}
\label{eq:HI_decomp}
 H_I({\bf q}_k)&=&\sum_{i=1}^L\alpha_i({\bf  q}_k) P_i\, ,
\end{eqnarray}
with $P_i\in\{\mathbb{1},X,Y,Z\}^{\otimes n}$ tensor products of Pauli matrices. 
The two point function can then be expressed as
\begin{equation}
 C(\tau,{\bf q}_k)=\sum_{i,i^\prime=1}^L\alpha_{i^\prime}({\bf q}_k)\alpha_i({\bf q}_k) s_{i^\prime,i}(\tau)\, ,
\end{equation}
where, for convenience, we have defined  the following matrix elements 
\begin{equation}
s_{i^\prime,i}(\tau)\equiv \langle\Psi_0\lvert V_{i^\prime i}(\tau)\rvert \Psi_0\rangle\; ,
\label{eq:siip}
\end{equation}
with $V_{i^\prime i}(\tau)=U^\dagger(\tau)P_{i^\prime}U(\tau)P_i$.
As explained in Ref.~\cite{Somma2002} the above quantity can be evaluated using the Hadamard test
for the unitary operators $V_{i^\prime i}$ and using gate identities we can write 
\begin{eqnarray}
\label{eq:had_test}
\Qcircuit @C=1em @R=1em {
	\lstick{\ket{0}} & \gate{H}&\ctrl{1}&\ctrl{1} &\ctrlo{1}&\ctrlo{1}&\qw &\meter \\
	\lstick{\ket{\Psi}} & \qw&\gate{P_{i}} & \gate{U(\tau)}&\gate{U(\tau)} &\gate{P_{i^\prime}}&\qw&\qw
}\, \raisebox{-1.2em}{,}
\end{eqnarray}
where we have in principle two controlled and two anti-controlled unitary operations.
Using the fact that a controlled and anti-controlled of the same unitary is equivalent to a single unitary acting only on the target qubits, as stated in Ref.~\cite{Somma2002}, we have
\begin{equation}
\label{eq:somma_resp}
\Qcircuit @C=1em @R=1em {
	\lstick{\ket{0}} & \gate{H}&\ctrl{1}&\qw & \ctrlo{1} &\qw&\meter \\
	\lstick{\ket{\Psi}}&\qw & \gate{P_i}&\gate{U(\tau)} & \gate{P_{i^\prime}} &\qw&\qw
}\, \raisebox{-1.2em}{.}
\end{equation}
The anti-controlled gate (i.e. a gate applied when the ancilla is in $\ket{0}$) may be expressed in terms of a controlled gate (i.e. a gate applied when the ancilla is in $\ket{1}$) used conjugation with the $X$ gate
\begin{equation}
\label{eq:somma_resp_fin}
\Qcircuit @C=1em @R=1em {
	\lstick{\ket{0}} & \gate{H}&\ctrl{1}&\gate{X} & \ctrl{1} &\gate{X}&\meter \\
	\lstick{\ket{\Psi}}&\qw & \gate{P_i}&\gate{U(\tau)} & \gate{P_{i^\prime}} &\qw&\qw
}\, \raisebox{-1.2em}{.}
\end{equation}
This modification is useful since it is often the case that gate libraries of current quantum devices have a controlled gate instead of an anti-controlled gate. In addition, the last $X$ gate can be avoided by appropriately redefining the measurement operators.

As first noted in Ref.~\cite{Somma2002}
the following quantum circuit 
allows us to compute $s_{i^\prime,i}(\tau)$ using only controlled Pauli operators and the application of one unitary transformation dependent on the time $t$, leading therefore to significantly shorter gate depths than a direct implementation of the circuit reported in Eq.~\eqref{eq:had_test}. Measuring the ancillary qubit in the basis corresponding to the eigenstates of the three Pauli matrices leads to
\begin{equation}
\langle X\rangle = \mathcal{R}e\left(s_{i^\prime i}(\tau)\right)\, ,\quad\langle Y\rangle =- \mathcal{I}m\left(s_{i^\prime i}(\tau)\right)\;,
\end{equation}
and it is easy to see that $\langle Z\rangle=0$. 
We notice that the total number of measurements necessary to obtain the above result with {\it statistical} precision $\epsilon$ is bounded by
\begin{eqnarray}
N\leq \frac{L^2}{\epsilon^2}\max_{k}\lvert\lvert \overline{H_I({\bf q}_k)}\rvert\rvert_2^4\leq\frac{L^4}{\epsilon^2}\max_k\max_i|\alpha_i({\bf q}_k)|^4\, ,
\end{eqnarray}
where we have defined, similarly to Ref.~\cite{roggero2020A}, the following expression
\begin{equation}
 \lvert\lvert \overline{H_I({\bf q}_k)} \rvert\rvert_q=\left (\sum_{i=1}^L\lvert \alpha_i({\bf q}_k)\rvert^q\right)^{1/q}\, , \quad \text{for}\, q\geq 1\; .
\end{equation}
A proof of this result may be found in Appendix~\ref{app:bounds_proof}.
The strong scaling with the number of terms can be mitigated in a number of different ways. On a fault-tolerant quantum device one can obtain $\mathcal{O}(L/\epsilon)$ using amplitude estimation~\cite{Brassard_2002,Knill2007} by trading the number of repetitions with a comparable increase in the gate depth. One can also completely remove the $L$ dependence from the number of measurements by directly applying the interacting Hamiltonian $H_I$, controlled by the ancila qubit in the circuit diagrams above. Alternately, one can implement the sum in Eq.~\eqref{eq:HI_decomp} using the "linear combination of unitaries (LCU)" algorithm~\cite{childs2012} and performing the quantum control on the PREPARE unitary instead (see also~\cite{Roggero2020b}). Of course, these two techniques can be combined together to obtain the benefits of both. 
The quantity we actually calculate on a quantum computer is 
\begin{eqnarray}
\label{eq:approx_c}
 \widetilde{C}(\tau,{\bf q}_k)&=& \bra{\Psi}V^\dagger(\tau)H_I({\bf q}_k)V(\tau)H_I({\bf q}_k)\ket{\Psi}\, ,
\end{eqnarray}
where $\ket{\Psi}$ is an approximation of the exact ground state $\ket{\Psi_0}$ with fidelity $F$ and $V(\tau)$ is a generic approximation of the time evolution which can be chosen such that
\begin{equation}
    \| U(\tau)-V(\tau)\|\leq \frac{\epsilon(\tau)}{2}\, .
\end{equation}
We can therefore bound the difference between the ideal response function and the approximate one as 
\begin{equation}
\label{eq:bound_error_c}
\lvert C(\tau,{\bf q}_k)-\widetilde{C}(\tau,{\bf q}_k)\rvert\!\leq\!\| \overline{H_I({\bf q}_k)}\|_1^2
\left(\!\epsilon(\tau)\!+\!\sqrt{1-F}\right),%\, ,
\end{equation}
where we have defined
\begin{equation}
    F=\lvert \bra{\Psi}\Psi_0\rangle\rvert^2\label{eq:F_bound}\, ,
\end{equation}
and a proof of this bound is provided in Appedix~\ref{app:trotter_bounds}.
\section{Lattice model mapped to qubits}
The model defined in Eq.~\eqref{eq:gen_model} can be mapped to qubits in various ways as described in Ref.~\cite{roggero2020A} (see also~\cite{DiMatteo2021} for an alternative based on the Gray code). In the following we will consider the case of two flavors and four lattice sites, with the first quantization mapping already employed in Ref.~\cite{roggero2020A} and that we report here for completeness
\begin{eqnarray}
\ket{1}=\ket{\upuparrows}\,,\quad \ket{3}=\ket{\downarrow\uparrow}\, ,\quad \ket{3}=\ket{\uparrow\downarrow}\, ,\quad \ket{4}=\ket{\downdownarrows} \nonumber\, .
\end{eqnarray}
We notice that with the mapping above we use $2$ qubits for each of the two particles to store their lattice site.
Using this mapping in the 4 qubit Hilbert space, the Hamiltonian operator from Eq.~\eqref{eq:gen_model} with the addition of the static contribution from Eq.~\eqref{eq:on_site_int} takes the form
\begin{equation}
\begin{split}
H&=8t+\frac{U}{2}-2t\sum_{k=1}^4X_k\\
&-\frac{U}{4}\left(Z_1Z_4+Z_2Z_3\right)-\frac{U}{4}\sum_{i<j<k}Z_iZ_jZ_k\, ,
\end{split}
\end{equation}
when mapped to qubits, the interacting Hamiltonian of interest from Eq.~\eqref{eq:int_ham} becomes
\begin{equation}
\begin{split}
H_I({\bf q}_k)~&= \sum_{f\in\{A,B\}} e_f \big[a_0({\bf q}_k)+a_1({\bf q}_k)Z_1\\
&~+a_2({\bf q}_k)Z_2+a_3({\bf q}_k)Z_3\big]\, ,
\end{split}    
\end{equation}
where explicit expressions of the coefficients $a_i({\bf q}_{k})$ are reported in Appendix~\ref{eq:amn}.
The initial state preparation follows the approach in Ref.~\cite{roggero2020A} and has a numerically exact fidelity  relative to the exact ground state of $96.2\%$ using the definition of fidelity reported in Eq.\eqref{eq:F_bound}.

\begin{figure}[t]
\centering
\includegraphics[width=0.5\textwidth]{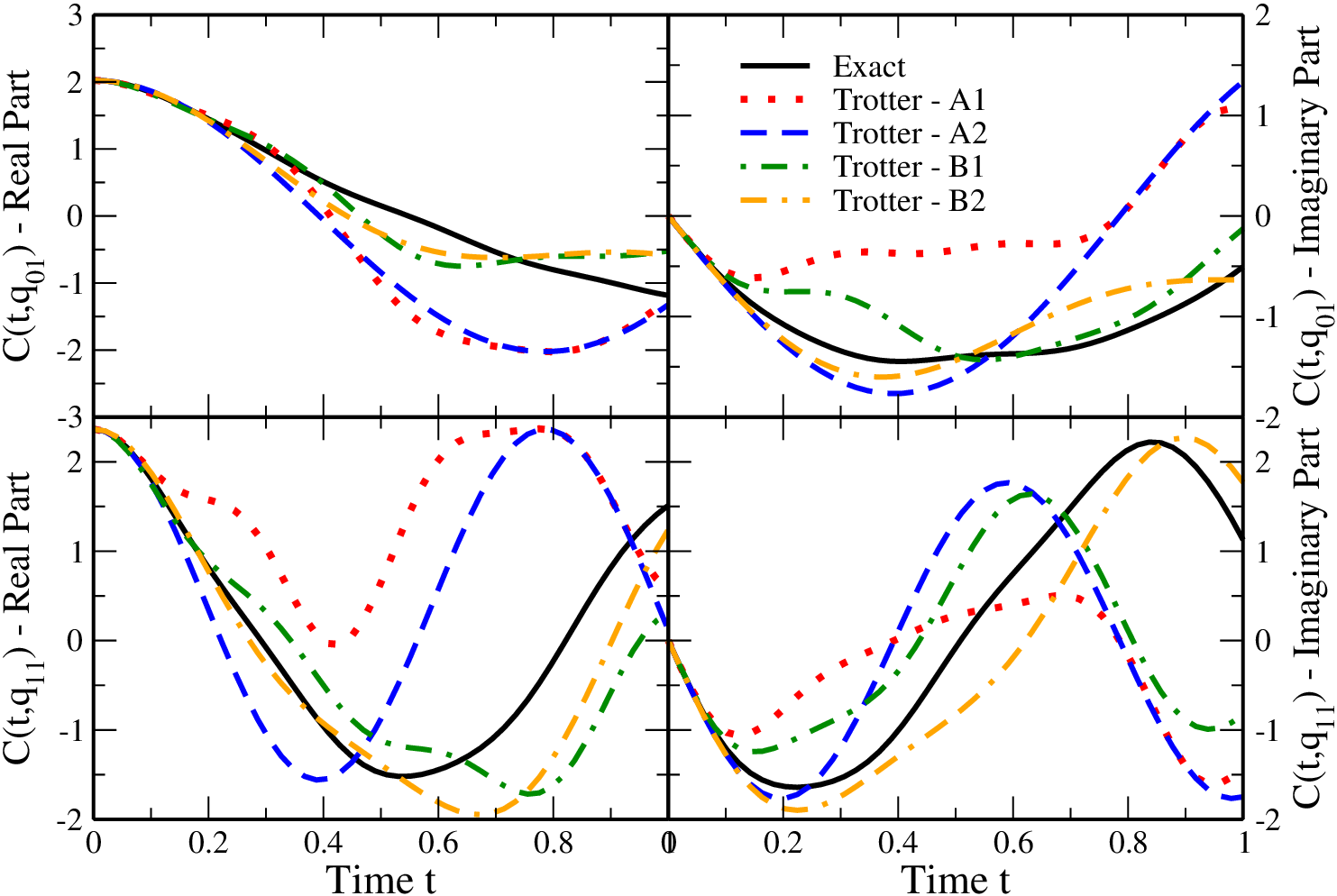}
\caption{Real and imaginary parts of the response function at different lattice sites using time orderings $A1$, $A2$, $B1$ and $B2$ compared against the numerically exact calculation, for the problem described in the text. The top panels are for ${\bf q}=(0,1)$ while the bottom panels are for ${\bf q}=(1,1)$.}
\label{fig:ex_corr}
\end{figure}

The calculation of Eq.~\eqref{eq:resp_real_time} requires the implementation of approximations of the time evolution operator leading to the approximate two-point function reported in Eq.~\eqref{eq:approx_c}. In this work we will use first order Trotter-Suzuki approximations. We define the following three Hamiltonians (using $bd$ as abbreviation for body)
\begin{equation}
\begin{split}
\label{eq:trotterizzations}
H_{A}^{(1bd)}&\equiv -2t\sum_{k=1}^4X_k\, ,\\
H_{A}^{(2bd)}&\equiv -\frac{U}{4}\left(Z_1Z_4+Z_2Z_3\right)\, ,\\
H^{(3bd)}&\equiv -\frac{U}{4}\sum_{i<j<k}Z_iZ_jZ_k\, .
\end{split}
\end{equation}
We recall that the time evolution for each of these Hamiltonians can be decomposed exactly into one- and two- qubit gates using fundamental circuit identities \cite{roggero2020A} (see also~\cite{Barenco1995} for general constructions). We can explicitly check that the total number of first-order Trotter-Suzuki splitting is four and we will group them in two categories called $A$ and $B$ and defined as following.
We will indicate as $A$ time orderings the following decompositions 
\begin{eqnarray}
U_{A1}(\tau)&=&e^{-iH_A^{(1bd)}\tau}\,e^{-iH_A^{(2bd)}\tau-iH^{(3bd)}\tau}\, ,\\
U_{A2}(\tau)&=&e^{-iH_A^{(2bd)}\tau-iH^{(3bd)}\tau}\, e^{-iH_A^{(1bd)}\tau}\, ,
\end{eqnarray}
and we note that the two-body and three-body part commute.
For the remaining two time orderings, it is convenient to define the following Hamiltonians
\begin{equation}
H_B^{(i,j)}=-2t(X_i+X_j)-\frac{U}{4} Z_iZ_j\, ,
\end{equation}
where $(i,j)\in\{(1,4),(2,3)\}$.
The real time evolution operators for the $B$ time orderings can be written as
\begin{eqnarray}
\label{eq:UB1_red}
U_{B1}(\tau)&=&e^{-i H_B^{(1,4)}\tau}e^{-i H_B^{(2,3)}\tau}e^{-iH^{(3bd)}\tau}\, ,\\
U_{B2}(\tau)&=&e^{-iH^{(3bd)}\tau}e^{-i H_B^{(1,4)}\tau}e^{-i H_B^{(2,3)}\tau}\, .
\end{eqnarray}
We notice that the commutator $[H^{(i,j)},H^{(k,l)}]$ for $i\neq j\neq k \neq l$ is vanishing.
We report in Fig.~\ref{fig:ex_corr} the comparison between the exact and the approximate time evolutions for the problem considered at different lattice sites for the full response functions.
We note that for small times all the time orderings seem to be quite close to the exact time evolution. The ordering $B2$  works better than the others for the imaginary parts of the response function for longer times. We also notice that sums rules are exact for the $B$-type propagators.
\begin{figure*}[th]
$$\Qcircuit @C=1em @R=1em {
	\lstick{\ket{+}} &\ctrl{3}&\qw&\gate{X} & \qw&\ctrl{1} &\meter \\
	\lstick{T_1} & \qw&\multigate{3}{e^{-iH^{(3bd)}\tau} } & \qw&\multigate{1}{e^{-iH_B^{(1,4)}\tau}}& \gate{Z}&\qw \\
	\lstick{T_4}&\qw&\ghost{e^{-iH^{(3bd)}\tau} }&\qw \qw&\ghost{e^{-iH_B^{(1,4)}\tau}}&\qw&\qw\\
	\lstick{T_3}&\gate{Z}&\ghost{e^{-iH^{(3bd)}\tau}}&\qw&\multigate{1}{e^{{e^{-iH_B^{(2,3)}\tau}}}}&\qw&\qw\\
	\lstick{T_2}&\qw&\ghost{e^{-iH^{(3bd)}\tau}}&\qw&\ghost{e^{-iH_B^{(2,3)}\tau}}&\qw&\qw\\
}\raisebox{-4.1 em}{,}\qquad\qquad\quad \Qcircuit @C=1em @R=1em {
	\lstick{\ket{+}} &\ctrl{3}&\qw&\gate{X} & \qw&\ctrl{1} &\meter \\
	\lstick{T_1} & \qw & \qw&\multigate{1}{e^{-iH_B^{(1,4)}\tau}}&\multigate{3}{e^{-iH^{(3bd)}\tau} }& \gate{Z}&\qw \\
	\lstick{T_4}&\qw&\qw \qw&\ghost{e^{-iH_B^{(1,4)}\tau}}&\ghost{e^{-iH^{(3bd)}\tau}}&\qw&\qw\\
	\lstick{T_3}&\gate{Z}&\qw&\multigate{1}{e^{{e^{-iH_B^{(2,3)}\tau}}}}&\ghost{e^{-iH^{(3bd)}\tau} }&\qw&\qw\\
	\lstick{T_2}&\qw&\qw&\ghost{e^{-iH_B^{(2,3)}\tau} }&\ghost{e^{-iH^{(3bd)}\tau} }&\qw&\qw\\
} \raisebox{-5.1 em}{,} $$
 \caption{Circuits for the calculation of the correlator $\langle Z_1(t)Z_3\rangle$ using the time evolutions $B1$ (left) and $B2$ (right). \label{fig:somma_resp}}
\end{figure*}

In order to implement the decomposition from Eq.~\eqref{eq:UB1_red} in the circuit shown in Eq.~\eqref{eq:somma_resp_fin} we use the the three-body operator derived in Ref.~\cite{roggero2020A} and reported for completeness in Appendix \ref{app:circuits}.
We report in Fig.~\ref{fig:somma_resp}, as an example, the two circuits needed to calculate the correlator $\langle Z_1(\tau)Z_3\rangle$ using the two different time evolutions $U_{B1}(\tau)$ and $U_{B2}(\tau)$.  
We can obtain similar expressions for the $A$ orderings.
The CNOT gate count for the various time orderings and correlators for one Trotter step is reported in Tab.~\ref{tab:cnot_counts}.
We notice here that the fact that term $B1$ is much more expensive than term $B2$ can be easily understood looking at Fig.~\ref{fig:somma_resp}. As it can be seen for the ordering $B2$ the diagonal three-body term commutes with the second controlled $Z$ gate and can therefore be removed, while for the ordering $B1$ this simplification is not possible.
\begin{table}[b]
	\begin{tabular}{|l|l|r|l|l|}
		\hline
		$\quad$&  {\rm A1} & {\rm A2} & {\rm B1} & {\rm B2}\\ \hline
		$\langle Z_1(\tau)Z_1\rangle$ &    $19$    &  $6$       &    $26$   &  $8$  \\ \hline
		$\langle Z_1(\tau)Z_3\rangle$& $25$ & $9$ & $28$ & $11$ \\ \hline
		$\langle Z_1(\tau)Z_2(\tau)Z_3Z_4\rangle$ &$25$ & $15$  & $28$ & $15$ \\ \hline
		$\langle Z_1(\tau)Z_2(\tau)Z_1Z_2\rangle$ &$30$ & $9$ & $29$ & $13$\\ \hline
	\end{tabular}
\caption{ CNOT gate count for the implementation of the two-point function calculation using different decompositions of the time evolution operator. Different rows correspond to different operator structures (see Appendix~\ref{eq:amn} for details).~\label{tab:cnot_counts}}

\end{table}

A more detailed example of the implementation of the above algorithm for the calculation of the two correlators $\langle Z_1(\tau) Z_3\rangle$ and $\langle Z_4(\tau)Z_3(\tau)Z_2Z_1\rangle$ is reported in Appendix~\ref{app:circuits}.

\section{Results}
We report here results for both the real and imaginary part of the approximate two point correlation functions $\widetilde{C}(\tau,{\bf q}_k)$ from Eq.~\eqref{eq:approx_c} obtained using the IBM five qubit machine Ourense~\cite{IBMQ_Ourense}. We used the IBM quantum programming language called Qiskit ~\cite{Qiskit21s} to implement the above circuits.
We applied to the results of the noisy quantum processing units error mitigation protocols developed in Refs.~\cite{roggero2020A}. In particular we first used readout error mitigation with associated error propagation, and then the zero-noise-extrapolation (ZNE) as described in Ref.~\cite{roggero2020A} (see also Refs.~\cite{kandala2017,Li2017,Dumitrescu2018,Endo2018} and the appendices of~\cite{Roggero2020b,Hall2021} for more details on our implementation).

\begin{figure*}
	\centering
	\includegraphics[width=0.5\textwidth]{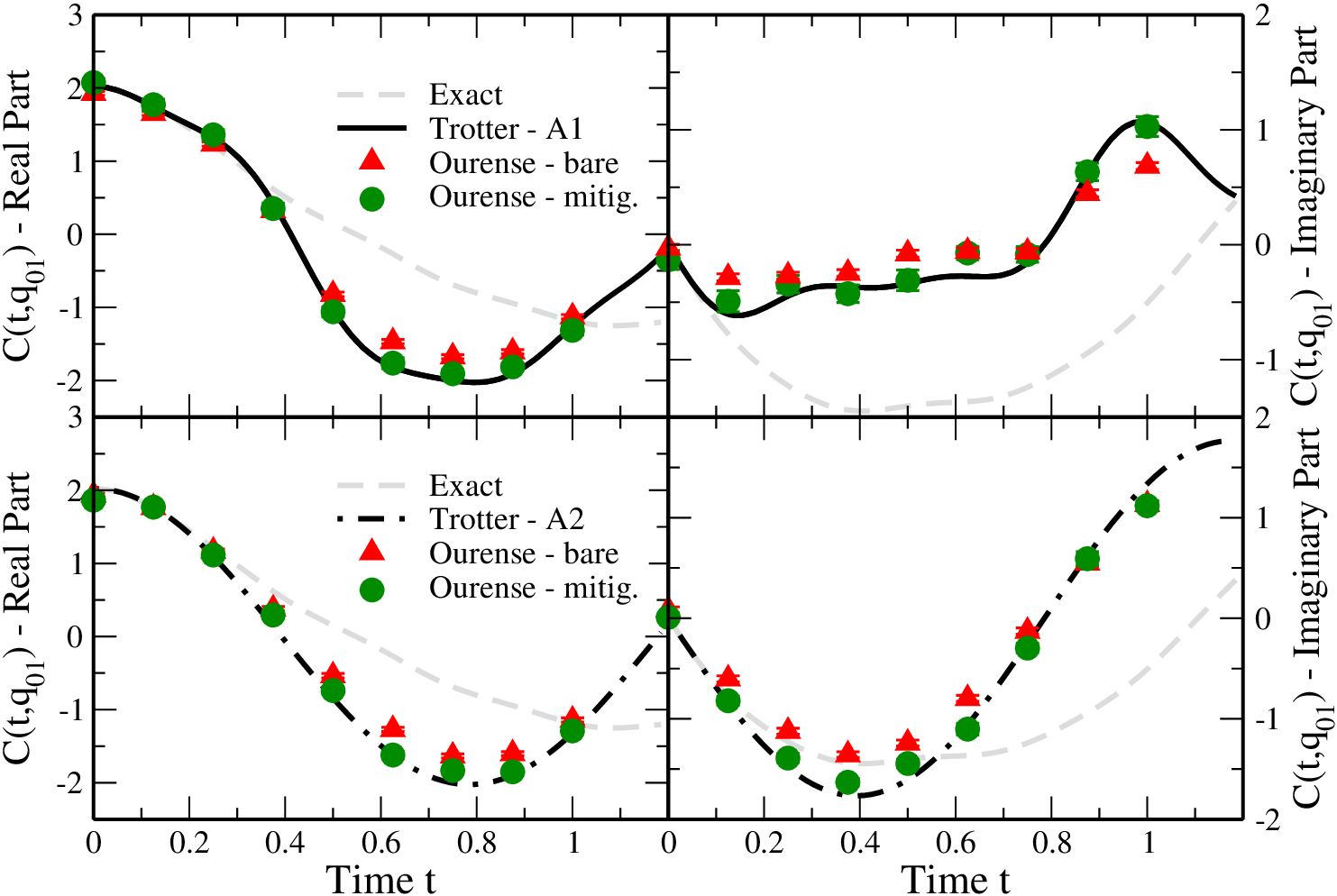}\includegraphics[width=0.5\textwidth]{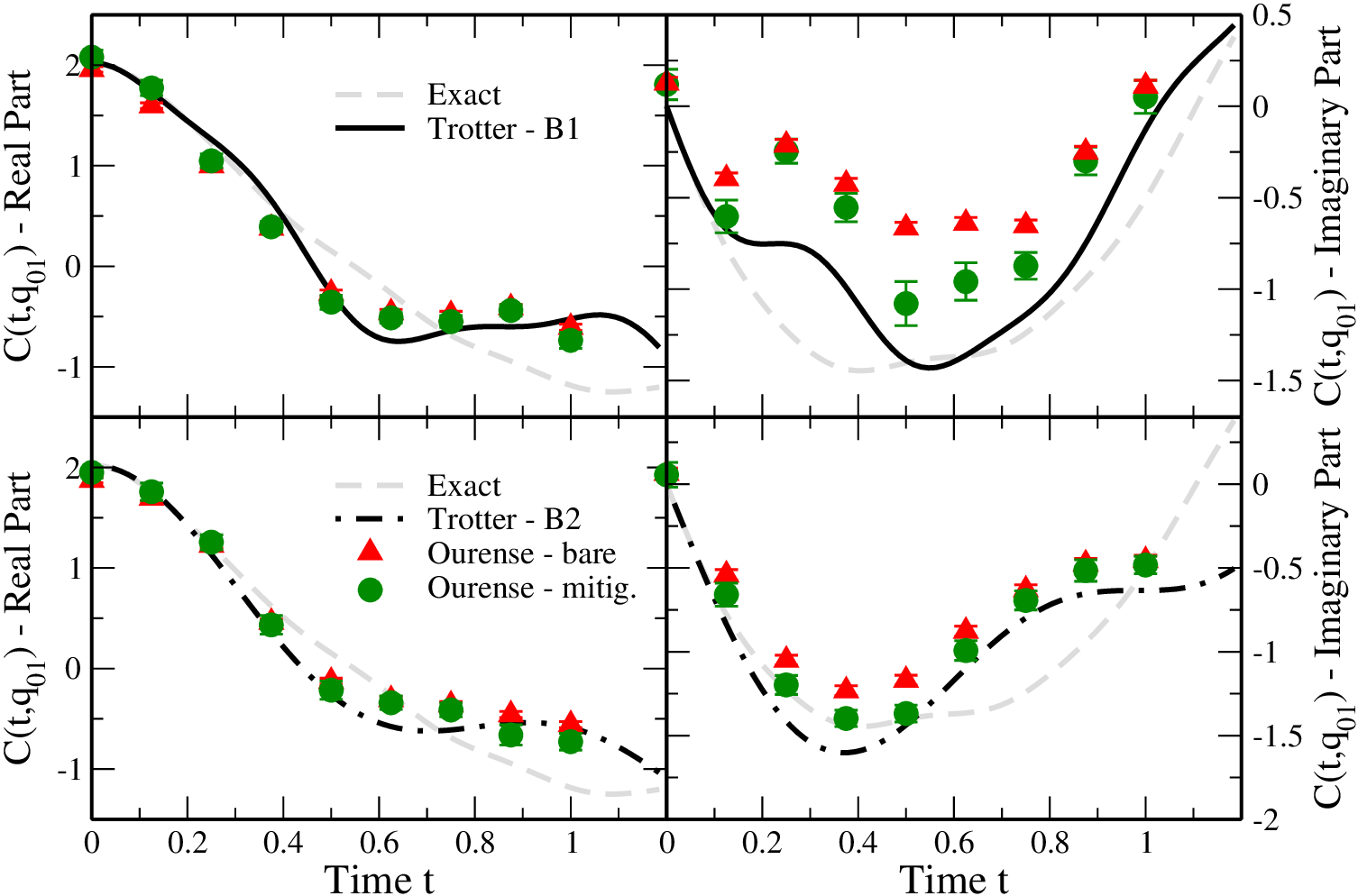}
	\caption{ Real and imaginary parts of the response function wave-vector ${\bf q}=(0,1)$ using time ordering $A1$, $A2$ (left panel) and $B1$, $B2$ (right panel) obtained from the five qubit machine Ourense~\cite{IBMQ_Ourense}. For the Trotterized time evolution, black lines, green and red dots denote the numerically exact result, the bare estimate from the QPU and the error mitigated results, respectively. The dashed grey line represents the exact dynamics with no time step errors.}
	\label{fig:all_corr01}
\end{figure*}

\begin{figure*}
	\centering
	\includegraphics[width=0.5\textwidth]{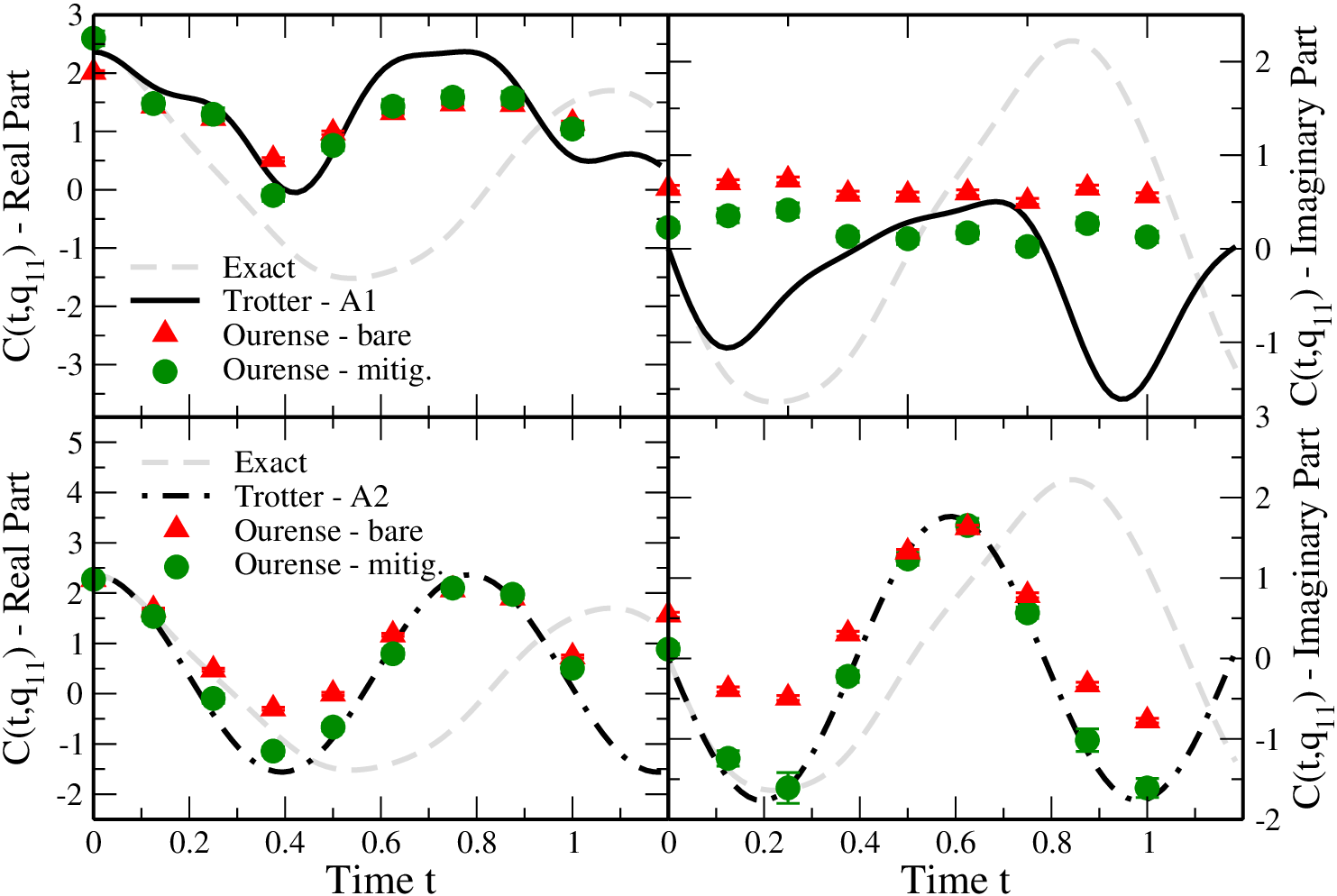}\includegraphics[width=0.5\textwidth]{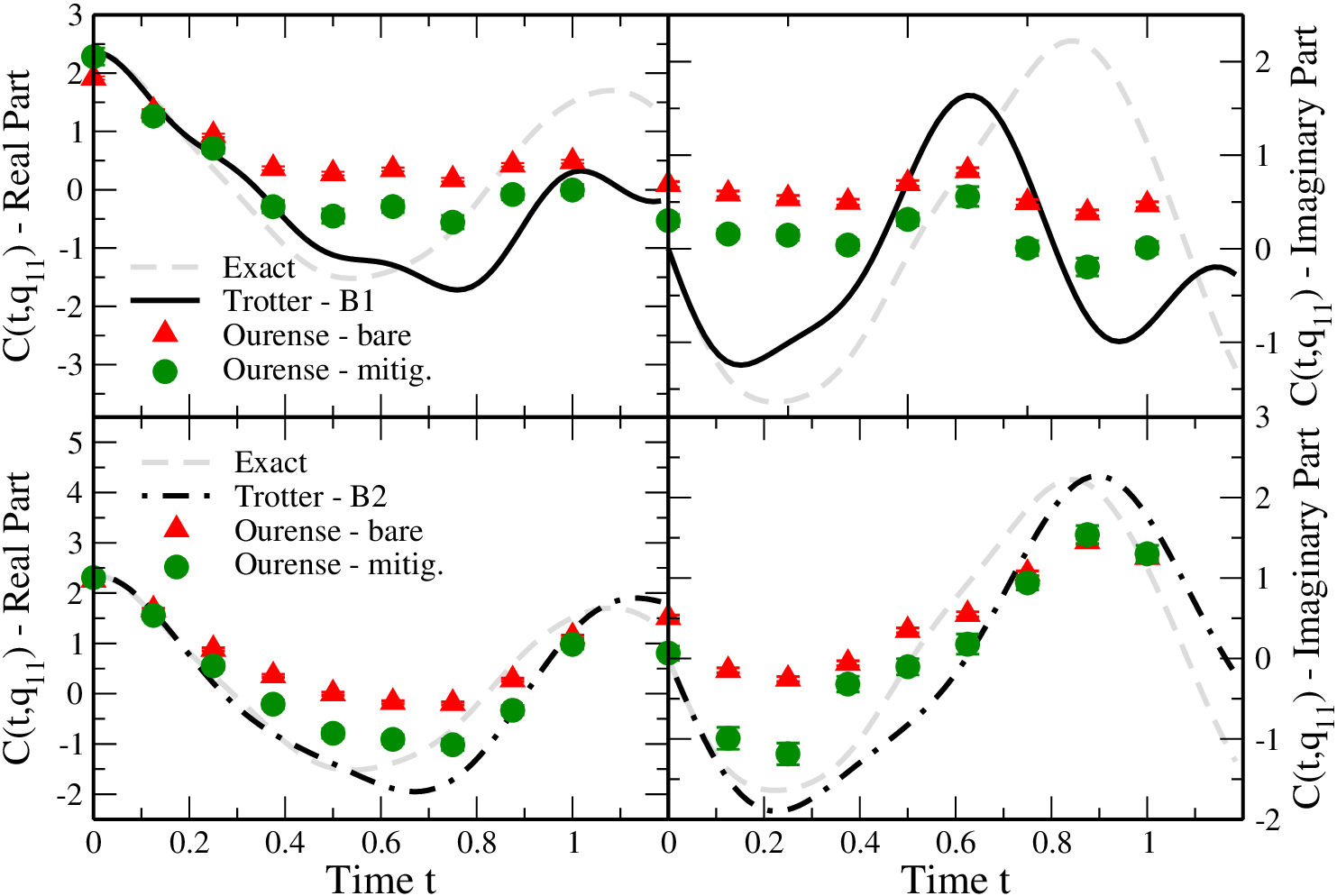}
	\caption{ Real and imaginary parts of the response function wave-vector ${\bf q}_1=(1,1)$ using time ordering $A1$, $A2$ (left panels) and $B1$, $B2$ (right panels) ran on the five qubit machine Ourense. Black lines, green and red dots denote respectively the expected result, the bare estimate from the QPU and the error mitigated results. The dashed grey line represents the exact dynamics with no time step errors.}
	\label{fig:all_corr11}
\end{figure*}
The final two point functions at two different values of the momentum transfer are shown in Figs.~\ref{fig:all_corr01}--\ref{fig:all_corr11}. 
We notice that for the real and imaginary part of the two point correlator at momentum transfer ${\bf q}=(0,1)$, the bare results on current quantum hardware deviate only slightly from the numerical results, and the applied error mitigation procedure brings the machine results very close to the numerical ones.
In particular we notice that the time ordering $A2$ and $B2$ lead to better agreement with the numerical calculations with respect to their counterparts $A1$ and $B1$. The higher infidelity could have been anticipated by noticing the different CNOT gate count (higher for the orderings $A1$ and $B1$ compared to the orderings $A2$ and $B2$, see Tab.~\ref{tab:cnot_counts}).
We also notice that the first order Trotter approximation used here is a good approximation for the exact time evolution for times around $0.1-0.2$ (see results in Fig.~\ref{fig:ex_corr}), however, in an actual calculation aimed at reproducing the exact time evolution higher order Trotter formulas should be used.
The calculation of the real response function at momentum transfer ${\bf q}=(1,1)$ shows more discrepancies with numerically exact results, and in particular for the imaginary part the mitigated hardware results are significantly different from the exact ones.
In order to asses the quality of the above runs, similarly to what has been done in Ref.\cite{Roggero2020b}, we introduce the following error metrics, the chi squared and the normalized sum of squared deviations (nssd), defined respectively as
\begin{eqnarray}
\label{eq:emetrics}
 \chi^2({\bf q}_k)&=&\sum_{l=1}^{N}\frac{\left[C^{(t)}(\tau_l,{\bf  q}_k)-C^{(e)}(\tau_l,{\bf q}_k)\right]^2}{[\Delta C(\tau_l,{\bf q}_k)]^2}\, ,\\
 {\rm nssd}(r,{\bf q}_k)&=&\sqrt{\frac{\sum_{l=1}^N\left[C^{(t)}(\tau_l,{\bf  q}_k)-C^{(e)}(\tau_l,{\bf q}_k)\right]^2}{\sum_{l=1}^N[r C^{(t)}(\tau_l,{\bf q}_k)]^2}}\, ,\nonumber\\
\end{eqnarray}
where $N$ is the number of time steps used, and $C^{(e)}(\tau_l,{\bf q}_k)$ and $C^{(t)}(\tau_l,{\bf q}_k)$ denote respectively the exact theoretical and the experimental results. These two error metrics serve two distinct purposes: the $\chi^2$ is used as an indication that the error estimation is appropriate and large values indicate a residual component of unaccounted systematic error, the $nssd$ metric instead is useful to understand the performance of the error mitigation routine in reproducing the centroid of the distribution of the observables.
In the following we use for ${\rm nssd}$ a value of $r=0.1$ that denotes a $10\%$ relative error.
We first discuss the real and imaginary time correlation function at momentum transfer ${\bf q}=(0,1)$ reported in Table ~\ref{tab:resp_01}. The value of the error mitigated $\chi^2$ is between $10$ and $50$ times smaller than the corresponding bare (unmitigated value). For the ${\rm nssd}$ metric instead the mitigated value is between $>1$ and $5$ times smaller than the unmitigated ones. This suggests that the error mitigation techniques in this case favors the improvement of the dispersion of the results. We also notice that the worst performing ordering, i.e. with corresponding higher error metrics, is $B1$ for both cases and that can be explained by the high gate depth of its implementation.
We can now discuss the quality metrics for the real and imaginary part of the correlator at momentum transfer ${\bf q}=(1,1)$.
While the error mitigated values of both $\chi^2$ and ${\rm nssd}$ experience a reduction with respect to the bare values similar to the previous case, the final mitigated numbers are much higher.
We notice that the calculation of the two-point function at momentum transfer ${\bf q}=(1,1)$ requires four controlled operations that lead to circuit depths in some cases slightly higher than the case where only two controlled operations were required.
The imaginary part, in this case, shows the highest values for both the error metrics used.
Similar results have been obtained using other five qubit machines such as Vigo. Also, these calculations were done over several months and the results did not change.
The ordering $A2$ seems overall to provide an imaginary correlation function closer to the numerically exact one and this is related to the low gate count of the associated circuits used for the calculation.
The second ordering that performs well for the calculation of the imaginary response is $B2$ (which corresponds to the circuits with the second smallest CNOT count).
We finally arrive at the remaining orderings $A1$ and $B1$, that have comparably high CNOT gate counts and give considerably worse results.
However, we notice that the gate count for circuits used to obtain the real and imaginary part differs only of a phase gate and therefore gate count alone cannot explain the fact that the real parts for orderings $A1$ and $B1$ show results much more in agreement with the exact calculations than their imaginary counterparts. 
It is possible that the final state produced by the implemented quantum circuit using gates with limited fidelity has a larger deviation in the $Y$ direction (which we use to measure the imaginary part of the correlator) than in the $X$ direction (used for the real part).
\begin{table*}
\begin{tabular}{|*2{p{10mm}|}*4{p{23mm}|}}
\hline
\cline{3-6}
&&$A1$&$A2$&$B1$&$B2$\\
\hline
$\chi^2$&bare&$514.14\, (472.05)$&$658.52\, (716.81)$&$343.41\,(2456.35)$&$233.65\, (643.87)$\\
\cline{2-6}
&mit.&$11.88\,(20.43)$&$64.82\,(51.19)$&$75.14\,(167.15)$&$30.80\,(60.96)$\\
\hline
nssd&bare&$1.36\,(4.28)$&$1.60\,(2.28)$&$1.71\, (5.48)$&$1.44\,(2.40)$\\
\cline{2-6}
&mit.&$0.39\,(1.96)$&$0.66\,(0.94)$&$1.55\,(3.67)$&$1.25\, (1.39)$\\
\hline
\end{tabular}
	\caption[30mm]{Quality metrics for the real and imaginary (in parenthesis) part of the response function with ${\bf q}=(0,1)$ using different time orderings.\label{tab:resp_01}}
\end{table*}

\begin{table*}
\begin{tabular}{|*2{p{10mm}|}*4{p{27mm}|}}
\hline
\cline{3-6}
&&$A1$&$A2$&$B1$&$B2$\\
\hline
$\chi^2$&bare&$2885.02\, (13422.84)$&$3879.64\, (5212.32)$&$11305.15\,(11390.61)$&$9958.04\, (9774.88)$\\
\cline{2-6}
&mit.&$144.77\,(1547.30)$&$99.97\,(14.79)$&$320.89\,(1694.40)$&$270.56\,(303.41)$\\
\hline
nssd&bare&$3.16\,(15.93)$&$4.17\,(5.82)$&$8.45\, (11.78)$&$7.03\,(7.05)$\\
\cline{2-6}
&mit.&$2.61\,(12.17)$&$1.62\,(0.96)$&$4.91\,(9.16)$&$3.50\, (4.27)$\\
\hline
\end{tabular}
	\caption[30mm]{Quality metrics for the real and imaginary (in parenthesis) part of the response function with ${\bf q}=(1,1)$ using different time orderings.\label{tab:resp_01_real}}
\end{table*}

\section{From the time domain to the frequency domain}
We consider a quantum circuit similar to the one displayed in Eq.~\eqref{eq:somma_resp_fin} but with an initial time evolution on the target state $\ket{\Psi}$. This leads to the calculation of  the following correlator depending on two time variables 
\begin{eqnarray}
 C(\tau,t)&=&\bra{\Psi} H_I(t)H_I(t+\tau)\ket{\Psi}
\end{eqnarray}
where $H_I$ denotes a generic interacting Hamiltonian (where for convenience we left the possible dependence from the momentum transfer implicit, since it is of no interest in the following discussion) in the Heisenberg representation evolved with $U(t)=e^{-iHt}$. We notice that the correlation function of the previous sections is recovered whenever $t=-\tau$. Inserting a complete set of eigenstates $\ket{n}$ of $H$ we get
\begin{eqnarray}
 C(\tau,t)&=&\sum_{n,k,m}\bra{\Psi}m\rangle \bra{m}H_I\ket{n}\bra{n}H_I\ket{k}\bra{k}\Psi\rangle\nonumber\\
 &\times& e^{i(E_n-E_k)\tau}e^{i(E_m-E_k)t}\, ,
\end{eqnarray}
which allows us to define the following frequency domain version of the correlator $C(\tau,t)$
\begin{eqnarray}
S(\omega)&\equiv& \int_{-\infty}^{+\infty} \frac{dt}{2\pi}\int_{-\infty}^{+\infty}  \frac{d\tau}{2\pi} e^{i\omega \tau} C(\tau,t)\\
 &=&\sum_{n,k,m}\bra{\Psi}m\rangle \bra{m}H_I\ket{n}\bra{n}H_I\ket{k}\bra{k}\Psi\rangle\nonumber\\
 &\times& \delta(E_n-E_k+\omega)\delta(E_m-E_k)\, .
\end{eqnarray}
Using the second delta function in the line above  we arrive at the following expression for the frequency domain response function
\begin{equation}
 S(\omega)=\sum_{n,m}\left\lvert\bra{\Psi}m\rangle\bra{m}H_I\ket{n}\right\rvert^2\delta(E_n-E_m+\omega)\, ,
\end{equation}
where we recall that the above derivation assumes that the delta function over the energies is the same delta function over the states and therefore the states are not degenerate in energy. 
We notice that in practice the  integrals over $t$ and $\tau$ are done  for finite time and the frequency domain response function reads as
\begin{equation}
 \widetilde{C}^{\rm finite}(\widetilde{\tau},T,\omega)=\frac{1}{4T\widetilde{\tau}}\int^{T}_{-T}dt\int_{-\widetilde{\tau}}^{\widetilde{\tau}}d\tau e^{i\omega \tau} C(\tau,T)\, .
\end{equation}
We notice that in order to achieve a resolution $\Delta \omega$ we have $T\sim 1/\Delta\omega$.
In order to perform the above integral we define a grid over times $T$ ($\tau$) with $(2N_T+1)\Delta t=2T$ ($(2N_\tau+1)\Delta \tau=2\tau$) time steps. We can then write
\begin{equation}
 \widetilde{C}(\widetilde{
\tau},T,\omega)\equiv \frac{\Delta\tau\Delta t}{4 T\widetilde{\tau}} \sum_{j=-N_\tau}^{N_\tau} \sum_{l=-N_t}^{N_t} e^{i\tau_j\omega} C(\tau_j,t_l)\, ,
\end{equation}
where we used $\tau_j=j\Delta\tau$ and similarly $t_l=l\Delta T$. We also notice that we have the following upper bound using the midpoint rule in Riemann sums 
\begin{eqnarray}
 \lvert \widetilde{C}^{\rm finite}(T,T,\omega)- \widetilde{C}(T,T,\omega) \rvert \leq \alpha\frac{T^3\widetilde{\tau}^3}{N_T^2N_{\widetilde{\tau}}^2}\, ,
\end{eqnarray}
where $\alpha=M_2/24$, with $M_2$ maximum of the absolute value of the second derivative of the integrand on the interval of interest.
Choosing $\Delta t=\Delta\tau=\Delta$ and $\widetilde{\tau}=T$ we obtain
\begin{eqnarray}
\widetilde{ C}(T,T,\omega)&=&\frac{\Delta^2}{4\, T^2}  \sum_{j=-N_t}^{N_t} \sum_{l=-N_t}^{N_t} e^{i\,j\Delta\,\omega} C(j\Delta,l\Delta)\, ,
\end{eqnarray}
and in order to reach a precision  $\Delta \omega$ in the response function in frequency domain we need to evaluate $N_t^2={\mathcal O}(1/\Delta\omega^2)$ correlators.
This is quadratically worse than the ${\mathcal O}(1/\Delta\omega)$ cost (either in measurements or gate depth) required to compute the response function in frequency space using expansions in orthogonal polynomials~\cite{Somma2019,roggero2020C} or using Quantum Phase Estimation~\cite{roggero2019}. This suggests that, in situations where the frequency response is the observable of interest and when only approximations to the nuclear ground-state with low fidelity are available, these latter methods should be preferred. Future work in this direction will help elucidate the tradeoffs between the two approaches. A preliminary discussion of the effects of the excited contamination in the initial state is reported in Appendix~\ref{app:Excited-state-cont}.

\section{Conclusions and outlook}
In this paper we performed a calculation of the real time two-point correlation function for a simple model of a triton using leading order pion-less  EFT as first presented in Ref.~\cite{roggero2019}. 
In particula, we adopted for our problem an algorithm first presented in Ref.~\cite{Somma2002}. We performed a theoretical error analysis for the general version of the problem at hand investigating the required number of measurements as a function of the statistical precision and the error on the implementation of the time evolution. We performed calculations using current available IBM quantum hardware (five qubit machines, with $T$ connectivity).
Results have been error mitigated using readout error mitigation and zero-noise-extrapolation and are in agreement with the exact ones for the momentum transfer ${\bf q}=(0,1)$ and for the real part at momentum transfer ${\bf q}=(1,1)$.
We noticed discrepancies with the exact results for two of the four time orderings for the error mitigated imaginary part of the response function at ${\bf q}=(1,1)$. We can conclude that for almost all cases $A2$ leads to better error metrics than the ordering $B2$, which can be understood looking at the CNOT gate count reported in Tab.~\ref{tab:cnot_counts}.
In the future we plan to explore the use of different error mitigation strategies and to perform calculations for a larger number of sites.
Finally, we have also shown that techniques to reconstruct the frequency response starting from real-time data coming from two-point correlation function can require up to a quadratic increase in the number of experiments when the initial state is not the exact ground-state as compared with methods that work directly in frequency space. Besides applications where real time information is directly required, real time approaches might still be used efficiently in practice when either an approximation to the initial many-body ground state can be obtained with high-fidelity.

\begin{acknowledgements}
We thank L. Cincio and Y. Subasi for useful discussions. 
This work was supported in part by the U.S. Department of Energy, Office of Science, Office of Nuclear Physics, Inqubator for Quantum Simulation (IQuS) under Award Number DOE (NP) Award DE-SC0020970, by the DOE HEP QuantISED grant KA2401032 and by the Institute for Nuclear Theory under U.S. Department of Energy grant No. DE-FG02-00ER41132.
This manuscript has been authored in part by Fermi Research Alliance, LLC under Contract No. DE-AC02-07CH11359 with the U.S. Department of Energy, Office of Science, Office of High Energy Physics.
This research used resources of the Oak Ridge Leadership Computing Facility, which is a DOE Office of Science User Facility supported under Contract DE-AC05-00OR22725.

\end{acknowledgements}
\bibliography{biblio}

\appendix
\section{Exicitation operator and response function expansion}
\label{eq:amn}
We can explicitly write the density excitation operator for one of the fermion species, denoted with the subscript $A$ as 
\begin{equation}
\rho_A({\bf q}_k)=e_A\sum_{i=1}^4e^{i{\bf q}_k\cdot{\bf x}_i}c^\dagger_{i,A}c_{i,A}\, ,
\end{equation}
where $e_A$ is the charge. The total density operator  for both fermion species is simply $\rho=\rho_A+\rho_B$.
Using the mapping reported in Ref.~\cite{roggero2020A} we can write the excitation operator for the specie $A$ at site $(m,n)$ as
\begin{equation}
\begin{split}
\rho_A({\bf q}_k)&=e_A\big[\rvert00\rangle\langle00\lvert+(-1)^m\rvert01\rangle\langle01\rvert\nonumber\\
&+(-1)^n\rvert10\rangle\langle10\lvert+(-1)^{m+n}\rvert11\rangle\langle11\lvert\big]\, .
\end{split}
\end{equation}
\iffalse
that can be rewritten as
\begin{equation}
\begin{split}
\rho_A({\bf q}_k)
&=e_A\bigg[\frac{(1+(-1)^m)(1+(-1)^n)}{4} \nonumber\\
&+ \frac{(1+(-1)^m)(1-(-1)^n)}{4} Z_1 \nonumber\\
&+ \frac{(1-(-1)^m)(1+(-1)^n)}{4} Z_2\nonumber\\
& + \frac{(1-(-1)^m)(1-(-1)^n)}{4} Z_1Z_2\bigg]\;.
\end{split}
\end{equation}
\fi
The expression of the total charge $\rho=\rho_A+\rho_B$ can be written as
\begin{equation}
\begin{split}
\rho({\bf q}_{0,0})&=e_A+e_B\\
\rho({\bf q}_{0,1})&=e_AZ_1+e_BZ_3\\
%\rho({\bf q}_{1,0})&=e_AZ_2+e_BZ_4\\
\rho({\bf q}_{1,1})&=e_AZ_1Z_2+e_BZ_3Z_4\;.
\end{split}    
\end{equation}
Therefore we can decompose the response function for site $(0,1)$ as
\begin{equation}
\begin{split}
C(\tau,{\bf q}_{0,1}) &= \langle U^\dagger(\tau) \rho({\bf q}_{0,1})U(\tau)\rho({\bf q}_{0,1})\rangle = e_A^2 \langle Z_1(\tau)Z_1\rangle\\
&+ e_Ae_B \langle Z_1(\tau)Z_3\rangle +e_Ae_B \langle Z_3(\tau)Z_1\rangle\\
&+e_B^2 \langle Z_3(\tau)Z_3\rangle\;,
\end{split}
\end{equation}
where $Z_k(\tau)=U^\dagger(\tau) Z_kU(\tau)$, and similarly for the site $(1,0)$. 
\iffalse
For site $(1,0)$ we get 
\begin{eqnarray}
C(\tau,{\bf q}_{1,0})&= e_A^2 \langle Z_1(\tau)Z_1\rangle
+ e_Ae_B \langle Z_1(\tau)Z_3\rangle\nonumber\\
& +e_Ae_B \langle Z_3(\tau)Z_1\rangle+e_B^2 \langle Z_3(\tau)Z_3\rangle\;,
\end{eqnarray}
\fi
For the site $(1,1)$ we have
\begin{eqnarray}
C(\tau,{\bf q}_{1,1})&=&e_A^2\langle Z_1(\tau)Z_2(\tau) Z_1Z_2\rangle\nonumber\\
&+&e_Ae_B\left[\langle Z_1(\tau)Z_2(\tau)Z_3Z_4\rangle+\langle Z_3(\tau)Z_4(\tau)Z_1Z_2\rangle\right]\nonumber\\
&+&e_B^2\langle Z_3(\tau)Z_4(\tau)Z_3Z_4\rangle\, .
\end{eqnarray}
We recall here that 
both the initial state we consider and the Hamiltonian are symmetric respect to the exchanges
\begin{equation}
\quad(1,4)\leftrightarrow(2,3)\; .
\end{equation}

\section{Circuits}
\label{app:circuits}

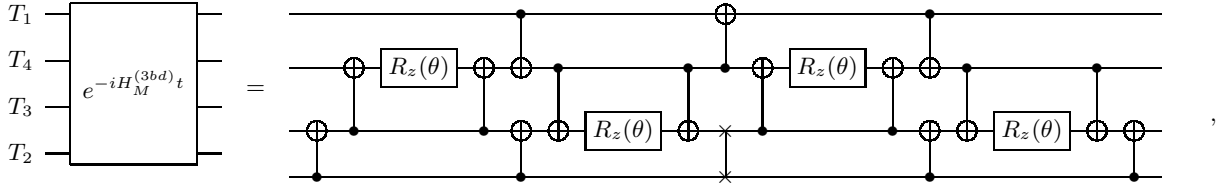
\begin{figure*}[th]
$$\Qcircuit @C=1em @R=1em {
	\lstick{T_1}&\multigate{3}{e^{-iH_M^{(3bd)}t}} &\qw\\
	\lstick{T_4}&\ghost{e^{-iH_M^{(3bd)}t} }&\qw\\
	\lstick{T_3}&\ghost{e^{-iH_M^{(3bd)}t}}&\qw\\
	\lstick{T_2}&\ghost{e^{-iH_M^{(3bd)}t}}&\qw\\
}\quad \raisebox{-3em}{=}\quad
\Qcircuit @C=0.7em @R=1em {
	 &\qw&\qw&\qw&\qw&\ctrl{1}&\qw&\qw&\qw&\targ&\qw&\qw&\qw&\ctrl{1}&\qw&\qw&\qw&\qw&\qw \\
	&\qw&\targ&\gate{R_z(\theta)}&\targ&\targ&\ctrl{1}&\qw&\ctrl{1}&\ctrl{-1}&\targ&\gate{R_z(\theta)}&\targ&\targ&\ctrl{1}&\qw&\ctrl{1}&\qw&\qw\\
	&\targ&\ctrl{-1}&\qw&\ctrl{-1}&\targ&\targ&\gate{R_z(\theta)}&\targ&\qswap&\ctrl{-1}&\qw&\ctrl{-1}&\targ&\targ&\gate{R_z(\theta)}&\targ&\targ&\qw\\
	&\ctrl{-1}&\qw&\qw&\qw&\ctrl{-1}&\qw&\qw&\qw&\qswap\qwx&\qw&\qw&\qw&\ctrl{-1}&\qw&\qw&\qw&\ctrl{-1}&\qw\\
}\qquad\raisebox{-4.1em}{,} $$
 \caption{Circuit decomposition for the three-body propagator for $e^{-iH_M^{(3bd)}t}$. The angle in the single qubit rotations is $\theta=\tau U/2$. \label{fig:three_body_prop}}
\end{figure*}

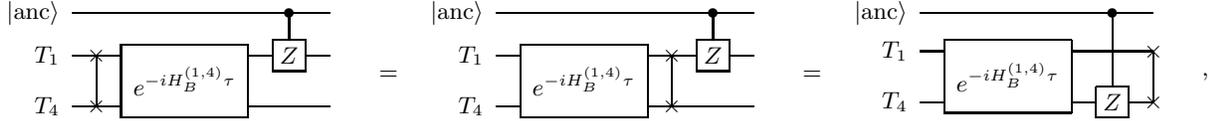
\begin{figure*}[th]
$$ \Qcircuit @C=1em @R=1em {
	\lstick{\ket{\rm anc}} &\qw&\qw&\ctrl{1} &\qw \\
	\lstick{T_1} &\qswap&\multigate{1}{e^{-iH_B^{(1,4)}\tau}}&\gate{Z} &\qw\\
	\lstick{T_4}&\qswap\qwx&\ghost{e^{-iH_B^{(1,4)}\tau}}&\qw&\qw\\
	}\qquad\raisebox{-2.5em}{=}\qquad\qquad \Qcircuit @C=1em @R=1em {
	\lstick{\ket{\rm anc}} &\qw&\qw&\ctrl{1} &\qw \\
	\lstick{T_1} &\multigate{1}{e^{-iH_B^{(1,4)}\tau}}&\qswap&\gate{Z} &\qw\\
	\lstick{T_4}&\ghost{e^{-iH_B^{(1,4)}\tau}}&\qswap\qwx&\qw&\qw\\
	}\qquad\raisebox{-2.5em}{=}\qquad\qquad \Qcircuit @C=1em @R=1em {
	\lstick{\ket{\rm anc}} &\qw&\ctrl{2} &\qw \\
	\lstick{T_1} &\multigate{1}{e^{-iH_B^{(1,4)}\tau}}&\qw &\qswap\\
	\lstick{T_4}&\ghost{e^{-iH_B^{(1,4)}\tau}}&\gate{Z}&\qswap\qwx\\
	}\qquad \raisebox{-2.3em}{,}$$
 \caption{Circuit identities used in the synthesis of the full circuit for pair correlators as described in the text \label{fig:gate_3q_swap}}
\end{figure*}

We preliminary discuss the three-body oracle reported in Fig.~\ref{fig:somma_resp} that can be expressed as follows
\begin{equation}
    \;\;\;\Qcircuit @C=1em @R=1em {
	\lstick{T_1} &\multigate{3}{e^{-iH^{(3bd)}t}}&\qw \\
	\lstick{T_4}&\ghost{e^{-iH^{(3bd)}t} }&\qw\\
	\lstick{T_3}&\ghost{e^{-iH^{(3bd)}t}}&\qw\\
	\lstick{T_2}&\ghost{e^{-iH^{(3bd)}t}}&\qw\\
}\;\;\;\; \raisebox{-3em}{=}\;\;\;\;%\qquad%\,\,
\Qcircuit @C=1em @R=1em {
	&\multigate{3}{e^{-iH_M^{(3bd)}t}}&\qswap &\qw\\
	&\ghost{e^{-iH_M^{(3bd)}t} }&\qswap\qwx&\qw\\
	&\ghost{e^{-iH_M^{(3bd)}t}}&\qswap&\qw\\
	&\ghost{e^{-iH_M^{(3bd)}t}}&\qswap\qwx&\qw\\
}\qquad\raisebox{-4.1em}{,}
\end{equation}
where the full decomposition of the oracle for $e^{-iH_M^{(3bd)}t}$, first derived in Ref.~\cite{roggero2020A}, can be found in Fig.~\ref{fig:three_body_prop}. %reads as
We describe now how we implemented the circuit reported in the right hand side of the $B1$ decomposition in Fig.~\ref{fig:somma_resp} (an similar construction applies to the $B2$ decomposition on the right). For ease of the present discussion, we temporarily remove the oracle gate that implements the time evolution $H_B^{(2,3)}$ given the fact that the second control is not acting on qubits $T_3$ and $T_2$. Now in order to simplify the circuit further we use the identity chain reported in Fig.~\ref{fig:gate_3q_swap} 
where $\ket{\rm anc}$ denotes a generic ancillary state, and in the first equality above we used the fact that $H_B^{(1,4)}=H_B^{(4,1)}$. The second equality in Fig.~\ref{fig:gate_3q_swap} follows from applying a swap gate to a controlled Z gate. Given the T connectivity of the machines used we want to apply the controlled gates over the same qubit. Starting from the final right hand side of Fig.~\ref{fig:gate_3q_swap} with the qubit $T_3$ added we can add an 
identity gate formed by two consecutive swap gates between qubits $T_4$ and $T_3$ 
%\begin{widetext}
\begin{eqnarray}
 \Qcircuit @C=1em @R=1em {
	\lstick{\ket{\rm anc}} &\qw&\qw&\ctrl{2} &\qw \\
	\lstick{T_1} &\qw &\qw&\qw&\qswap\\
	\lstick{T_4}&\qswap&\qswap&\gate{Z}&\qswap\qwx\\
	\lstick{T_3}&\qswap\qwx &\qswap\qwx &\qw&\qw\\
	}\quad\raisebox{-2.3em}{=}\qquad\quad \Qcircuit @C=1em @R=1em {
	\lstick{\ket{\rm anc}} &\qw&\ctrl{3} &\qw&\qw \\
	\lstick{T_1} &\qw &\qw&\qw&\qswap\\
	\lstick{T_4}&\qswap&\qw&\qswap&\qswap\qwx\\
	\lstick{T_3}&\qswap\qwx &\gate{Z} &\qswap\qwx&\qw\\
	}\qquad\raisebox{-2.3 em}{,}
	\label{eq:gate_3q_swap}
\end{eqnarray}
%\end{widetext}
and we can avoid executing the last two swaps between qubits $T_1, T_4$ and qubits $T_4,T_3$.

We report here one of the circuits used to calculate the correlator $\langle Z_4(\tau)Z_3(\tau)Z_2Z_1\rangle$.The implementation using the identities just described for the above mentioned correlator reads as
%\begin{widetext}
\begin{equation}
\label{eq:somma_resp2}
\Qcircuit @C=0.4em @R=1em {
	\lstick{\ket{+}} &\ctrl{2}&\qw&\gate{X} &\qw&\ctrl{2} &\gate{H}&\meter \\
	\lstick{T_1}& \qw&\multigate{3}{e^{-iH_M^{(3bd)}\tau} } & \multigate{1}{e^{-iH_B^{(1,4)}\tau}}&\qw&\qw&\qw&\qw \\
	\lstick{T_4}&\gate{Z}&\ghost{e^{-iH_M^{(3bd)}\tau} }&\ghost{e^{-iH_B^{(1,4)}\tau}}&\qw&\gate{Z}&\qw&\qw\\
	\lstick{T_3}&\gate{Z}\qwx&\ghost{e^{-iH_M^{(3bd)}\tau}}&\multigate{1}{e^{-iH_B^{(2,3)}\tau}}&\qw&\gate{Z}\qwx&\qw&\qw\\
	\lstick{T_2}&\qw&\ghost{e^{-iH_M^{(3bd)}\tau}}&\ghost{e^{-iH_B^{(2,3)}\tau}}&\qw&\qw&\qw&\qw\\
}\, \raisebox{-4.0em}{,}
\end{equation}
%\end{widetext}
where we notice that the last two swaps between qubits $1,4$ and $3,2$, located after the two remaining controlled Z are not reported in the circuit since they will not affect the measurement result. The above circuit can be reduced to the following form using circuit identities for sequences of Hadamard and controlled Z gates (see e.g Ref~\cite{Barenco1995})
%\begin{widetext}
\begin{equation}
\label{eq:somma_resp3}
\Qcircuit @C=0.4em @R=1em {
	\lstick{\ket{0}} & \qw&\targ&\qw&\qw&\gate{Z} &\qw&\targ &\meter \\
	\lstick{T_1}&\qw & \qw&\qw&\multigate{3}{e^{-iH_M^{(3bd)}\tau} } & \multigate{1}{e^{-iH_B^{(1,4)}\tau}}&\qw&\qw&\qw \\
	\lstick{T_4}&\targ&\ctrl{-2}&\targ&\ghost{e^{-iH_M^{(3bd)}\tau} }&\ghost{e^{-iH^{(1,4)}\tau}}&\targ&\ctrl{-2}&\qw\\
	\lstick{T_3}&\ctrl{-1}&\qw&\ctrl{-1}&\ghost{e^{-iH^{(3bd)}\tau}}&\multigate{1}{e^{-iH_B^{(2,3)}\tau}}&\ctrl{-1}&\qw&\qw\\
	\lstick{T_2}&\qw&\qw&\qw&\ghost{e^{-iH_M^{(3bd)}\tau}}&\ghost{e^{-iH_B^{(2,3)}\tau}}&\qw&\qw&\qw\\
}\, \raisebox{-3.7em}{.}
\end{equation}
%\end{widetext}

\section{Error bounds}
We quantify the total error given by the difference of the estimator and the Trotter error as $\widetilde{\epsilon}(t)$ and we require that the Trotter and the statistical  error to be less than $\widetilde{\epsilon}(t)/2=\epsilon(t)$.
In the following we analyze the two errors separately. 
\subsection{Measurement bounds}
\label{app:bounds_proof}
We estimate here the number of measurements needed to achieve the desired precision $\epsilon$ over the calculation of the real time response function $C(\tau,{\bf q}_k)$ .
We preliminary define the following estimator of the matrix element defined in the text in Eq.~\eqref{eq:siip}
\begin{eqnarray}
\overline{s}_{i^\prime,i}(\tau)=\frac{\sum_j s^j_{i^\prime,i}(\tau)}{M_{i^\prime,i}}\, ,
\end{eqnarray}
where $ s^j_{i^\prime,i}(\tau)$ denotes the outcomes of  $M_{i^\prime, i}$ measurements for a fixed value  of the time step $\tau$.
The estimator of the response function $C(\tau,{\bf q}_k)$ can be written as
\begin{eqnarray}
\overline{C}(\tau,{\bf q}_k)&=&\sum_{i,i^\prime=1}^L\,\alpha_{i^\prime}({\bf q}_k)\alpha_i({\bf q}_k)\overline{s}_{i^\prime,i}(\tau)\, ,\nonumber\\
\end{eqnarray}
where $L$ the number of terms in the interaction Hamiltonian is simply equal to the product of the number of species with nonzero coupling to the charge operator. The corresponding variance of the estimator reads as
\begin{eqnarray}
{\rm var}\left[\overline{C}(\tau,{\bf q}_{k})\right]&=&\sum_{i,i^\prime=1}^L\alpha^2_{i^\prime}({\bf q}_{k})\alpha^2_i({\bf q}_{k})\frac{1-\lvert\overline{s}_{i^\prime,i}(\tau)\rvert^2}{M_{i^\prime,i}}\, .\nonumber\\
&&\leq \sum_{i,i^\prime=1}^L\alpha^2_{i^\prime}({\bf q}_{k})\alpha^2_i({\bf q}_{k})\frac{1}{M_{i^\prime,i}}\, ,\label{eq:Cbar}
\end{eqnarray}
where the expression in the first line comes from the fact that the average of $s^2$ is the identity since (c.f. definition of $s_{i^\prime,i}(\tau)$ in Eq.~\eqref{eq:siip})
\begin{equation}
 \langle\psi_0\lvert V^\dagger_{i^\prime i}(\tau)V_{i^\prime i}(\tau)\rvert \psi_0\rangle = 1\, .
\end{equation}
We now require the variance in Eq.~\eqref{eq:Cbar} to be equal to $\epsilon^2$ and we consider the case for which $\forall i,i^\prime\,\, M_{i^\prime,i}=M=N/L^2$ where $N$ is the total number of measurements done to estimate $\overline{C}(\tau,{\bf q}_k)$. We arrive at
\begin{eqnarray}
N&\leq& \frac{L^2}{\epsilon^2}{\rm max}_{k}\left[\sum_{i=1}^L\alpha^2_i({\bf q}_k)\right]^2\, .
\label{eq:first_bound}
\end{eqnarray}
\subsection{ Trotter and state error bounds}
\label{app:trotter_bounds}
We notice that the quantity that we want to calculate is $s_{k^\prime k}(\tau)$ reported in Eq.~\eqref{eq:siip} and the actual implementation contains two errors one coming from the initial state  preparation and the other coming from the approximate implementation of the time evolution operator $V(\tau)$.
We preliminary define the difference 
\begin{eqnarray}
 \Delta_{k^\prime k}(\tau)&=&Tr\left[ V^\dagger (\tau) P_{k^\prime}V(\tau)P_k\Pi\right]\nonumber\\
 &-&Tr\left[U^\dagger(\tau)P_{k^\prime}U(\tau)P_k\Pi_0\right]\, ,
\end{eqnarray}
where we have defined $\Pi_0=\ket{\Psi_0}\bra{\Psi_0}$ and $\Pi=\ket{\Psi}\bra{\Psi}$. We can therefore write
\begin{eqnarray}
 \Delta_{k^\prime k}(\tau)&=&\delta^{\rm Trott.}_{k^\prime k}(\tau)+\delta_{k^\prime k}^{\rm state}(\tau)\, ,
\end{eqnarray}
where we have defined
\begin{eqnarray}
 \delta^{\rm Trott.}_{k^\prime k}(\tau)&\equiv& Tr\left[ \left( V^\dagger (\tau) P_{k^\prime}V(\tau)-U^\dagger(\tau)P_{k^\prime}U(\tau)\right)P_k\Pi_0\right]\nonumber\\
\delta_{k^\prime k}^{\rm state}(\tau)&\equiv&Tr\left[ V^\dagger (\tau) P_{k^\prime}V(\tau)P_k(\Pi-\Pi_0)\right]\, .
\end{eqnarray}
we notice that 
\begin{eqnarray}
\delta^{\rm Trott.}_{k^\prime k}(\tau)&=&\bra{\Psi_0} (V^\dagger(\tau)-U^\dagger(\tau)+U^\dagger(\tau))P_{k^\prime}V(\tau)P_k\ket{\Psi_0}\nonumber\\
&&-\bra{\Psi_0}U^\dagger(\tau)P_{k^\prime}(U(\tau)-V(\tau)+V(\tau))P_k\ket{\Psi_0}\nonumber\\
&=&\bra{\Psi_0} (V^\dagger(\tau)-U^{\dagger}(\tau))P_k^{\prime}V(\tau)P_k\ket{\Psi_0}\nonumber\\
&&-\bra{\Psi_0}U^\dagger(\tau)P_{k^\prime}(U(\tau)-V(\tau))P_k\ket{\Psi_0}\, ,
\end{eqnarray}
taking the norm of the difference of the above quantity
\begin{eqnarray}
\lvert \delta^{\rm Trott.}_{k^\prime k}(\tau)\rvert&\leq& \lvert\lvert V^\dagger(\tau)-U^\dagger(\tau)\rvert\rvert\,\,\lvert\lvert  P_{k^\prime}V(\tau)P_{k}\rvert\rvert \nonumber\\
&+&\lvert\lvert V(\tau)-U(\tau)\rvert\rvert\,\, \lvert\lvert P_k U^\dagger(\tau)P_{k^\prime}\rvert\rvert\nonumber\\
&=&2\lvert\lvert V(\tau)-U(\tau) \rvert\rvert \, ,
\end{eqnarray}
where we used the fact that the norm of the product of three unitaries is the unity.
An extensive discussion for the case of Trotterizzation and qubitization is reported in Ref.~\cite{roggero2020A}.
We finally compute the difference $\delta_{k^\prime k}^{\rm state}(\tau)$ and using Eq.~(B8) of ref.~\cite{Roggero2020b} we arrive at 
\begin{eqnarray}
\lvert \delta_{k^\prime k}^{\rm state}(\tau) \rvert \leq \frac{1}{2}Tr \lvert \Pi-\Pi_0\rvert\leq \sqrt{1-F}\, ,
\end{eqnarray}
where $F$ is the fidelity defined in Eq.~\eqref{eq:F_bound}.
\subsection{Excited state contamination in the initial state}
\label{app:Excited-state-cont}
The correlator that we want to calculate is the following
\begin{eqnarray}
 C(\tau,{\bf q}_k)&=&e^{iE_0\tau}\bra{\Psi_0}H_I({\bf q}_k)U(\tau)H_I({\bf q}_k)\ket{\Psi_0}\,, \nonumber\\
\end{eqnarray}
where $\ket{\Psi_0}$ denotes the exact ground state and  $U(\tau)=e^{-iH\tau}$ the exact time evolution.
We observe that using a set of eigenstates of $H$, denoted by $\ket{i}$, where $\ket{i=0}=\ket{\Psi_0}$ we can write
\begin{equation}
    H_I({\bf q}_k)=\sum_{i,j}a_{i,j}({\bf q}_k)\ket{i}\bra{j}\, ,
\end{equation}
where $a_{i,j}({\bf q}_k)=\bra{i}H_I({\bf q}_k)\ket{j}$. We, therefore, have
\begin{eqnarray}
  C(\tau,{\bf q}_k)&=&e^{iE_0\tau}\sum_l e^{-iE_l\tau}a_{0,l}({\bf q}_k)a_{l,0}({\bf q}_k)\, .
\end{eqnarray}
The approximate ground state can be expressed as 
\begin{equation}
\ket{\Psi}=c_0\ket{\Psi_0}+\sum_{i=1}^K c_i\ket{i}\, ,
\end{equation}
and the approximate correlator is
\begin{eqnarray}
 C_{\Psi}(\tau,{\bf q}_k)&=&\sum_{i,j,m}a_{i,j}({\bf q}_k)a_{j,m}({\bf q}_k)e^{i(E_i-E_j)\tau}\nonumber\\
 &&\bra{\Psi}i\rangle\bra{m}\Psi\rangle\, .
\end{eqnarray}
We can now consider the case of imaginary time evolution in which the operator $U(\tau)$ is replaced by the operator $U_E(\tau_E)=e^{-H\tau_E}$.
The euclidean version of the real time correlator is
\begin{eqnarray}
 C^{\rm E}(\tau_E,{\bf q}_k)&=&e^{-E_0\tau_E}\sum_{l}e^{-E_l\tau_E} a_{0,l}({\bf q}_k) a_{l,0}({\bf q}_k)\, ,\nonumber\\
\end{eqnarray}
and  expanding we obtain
\begin{eqnarray}
 C^{\rm E}(\tau_E,{\bf q}_k)&=&e^{-2E_0\tau_E}a_{0,0}({\bf q}_k) a_{0,0}({\bf q}_k)\nonumber\\
 &+&e^{-E_0\tau_E}\sum_{l\neq 0}e^{-E_l\tau_E} a_{0,l}({\bf q}_k) a_{l,0}({\bf q}_k)\, ,\nonumber\\
\end{eqnarray}
where the second term is subleading compared to the first term (if $E_0<E_l$ $\forall l\neq 0$).
The approximate correlator in euclidean time is
\begin{eqnarray}
 C^{\rm E}_{\Psi}(\tau_E,{\bf q}_k)&=&\sum_{i,j,m}a_{i,j}({\bf q}_k)a_{j,m}({\bf q}_k)e^{-(E_i+E_j)\tau_E}c_i^\star c_m \,.\nonumber\\
\end{eqnarray}
The exact euclidean correlator has the spectral decomposition
\begin{eqnarray}
 C^{\rm E}_{\Psi}(\tau_E,{\bf q}_k)&=&\sum_{j,m} a_{0,j}({\bf q}_k)a_{j,m}({\bf q}_k)e^{-(E_0+E_j)\tau_E}c_0^\star c_m \nonumber\\
 &&\nonumber\\
 &+&\sum_{i\neq 0,j,m}a_{i,j}({\bf q}_k)a_{j,m}({\bf q}_k)e^{-(E_i+E_j)\tau_E}c_i^\star c_m\, ,\nonumber\\
\end{eqnarray}
where the leading terms,  
proportional to $e^{-2E_0\tau_E}$, are 
\begin{eqnarray}
\label{eq:cexp_final}
C^{\rm E}_{\Psi}(\tau_E,{\bf q}_k) &=&\lvert c_0\rvert^2 a_{0,0}({\bf q}_k)a_{0,0}({\bf q}_k)e^{-2E_0\tau_E}\nonumber\\
 &+&\sum_{m\neq 0}a_{0,0}({\bf q}_k)a_{0,m}({\bf q}_k)e^{-2E_0\tau_E} c_0^\star c_m \nonumber\\
 &+&\cdots \,.\nonumber\\
\end{eqnarray}
Terms represented by $\cdots$ get contributions from the excited states that are exponentially suppressed by the respective mass gaps. However, the second term present in Eq.~\eqref{eq:cexp_final} above, coming from the excited state contamination in the initial state $\ket{\Psi}$, is only suppressed by $c_m /c_0$ and not exponentially. It is therefore important to optimize the fidelity of $\ket{\Psi}$, ie, increase the overlap with $\ket{\Psi_0}$, in order to reduce the systematic bias in the results.
\typeout{}

\end{document}